\documentclass[hyper,a4paper]{JHEP3}
\usepackage{graphicx}
\usepackage{epsfig}
\usepackage{dcolumn}
\usepackage{bm}
\usepackage{url}
\usepackage{multirow}

\usepackage{cite}
           
\newcommand{\comment}[1]{}
\newcommand{\newc}{\newcommand}

\def\thetab0{\theta_{B_0}}

\def\r2{\sqrt 2}
\def\beq{\begin{equation}}
\def\eeq{\end{equation}}
\def\bea{\begin{eqnarray}}
\def\eea{\end{eqnarray}}
\def\nn{\nonumber}
\def\wt{\widetilde}
\def\sinW2{\sin^2\theta_W}

\def\mz2{M_{z}^2}
\def\c2b{\cos 2\beta}

\def\mz{M_Z}

\def\bN{\mathbf N}
\def\bU{\mathbf U}
\def\bV{\mathbf V}
\def\bG{\mathbf G}
\def\bGpm{\mathbf G_{\pm}}

\def\sec2w{sec^2\theta_W}

\def\gmin2{(g-2)_\mu}

\catcode`\@=11 

\def\lsim{\mathrel{\mathpalette\@versim<}}
\def\gsim{\mathrel{\mathpalette\@versim>}}
\def\@versim#1#2{\vcenter{\offinterlineskipresembel
    \ialign{$\m@th#1\hfil##\hfil$\crcr#2\crcr\sim\crcr } }}

\newcommand{\ntrl}[1]{\widetilde{\chi}^0_#1}
\newcommand{\chpm}[1]{\widetilde{\chi}^{\pm}_#1}
\newcommand{\selL}{\widetilde{e}_L}
\newcommand{\smuL}{\widetilde{\mu}_L}
\newcommand{\selR}{\widetilde{e}_R}
\newcommand{\smuR}{\widetilde{\mu}_R}
\newcommand{\stau}[1]{\widetilde{\tau}_#1}

\newc{\ra}{\rightarrow}
\newc{\s}{\smallskip}
\newc{\non}{\noindent}

\def \chonep{{\wt\chi_1}^{+}}
\def \chonem{{\wt\chi_1^-}}
\def \chonep2{{\wt\chi_2^+}}
\def \chonem2{{\wt\chi_2^-}}

\def\mET{E_T \hspace{-1.2em}/\;\:}

\title
{Exploring novel correlations in trilepton channels
at the LHC for the minimal supersymmetric inverse seesaw model}
\author{Subhadeep Mondal,$^a$ ~Sanjoy Biswas,$^{b}$ ~Pradipta Ghosh$^c$ 
and Sourov Roy$^a$\\
$^a$Department of Theoretical Physics,\\
        Indian Association for the Cultivation of Science, \\
        2A $\&$ 2B Raja S.C. Mullick Road, Kolkata 700032, India\\

$^b$INFN, Sezione di Roma, Dipartimento di Fisica, Universit\`{a} di Roma La Sapienza\\
        Piazzale Aldo Moro 2;  I-00185 Rome, Italy\\

$^c$Departamento de F\'{\i}sica Te\'{o}rica UAM
        and Instituto de F\'{\i}sica Te\'{o}rica UAM/CSIC,\\
        Universidad Aut\'{o}noma de Madrid (UAM), Cantoblanco,
        28049 Madrid, Spain\\

E-mails:\email{tpsm2@iacs.res.in, Sanjoy.Biswas@roma1.infn.it, 
pradipta.ghosh@uam.es,  tpsr@iacs.res.in}}

\abstract{We investigate signatures of the minimal supersymmetric inverse seesaw model at 
the large hadron collider (LHC) with three isolated leptons and large missing energy ($3\ell
+ \mET~$ or $2\ell + 1\tau + \mET$, with $\ell=e,\mu$) in the final state.
This signal has its origin in the decay of chargino-neutralino $(\chpm1\ntrl2)$ 
pair, produced in $pp$ collisions. The two body decays of the lighter
chargino into a charged lepton and a singlet sneutrino has a characteristic 
decay pattern, which is correlated with the observed large atmospheric neutrino mixing 
angle. This correlation is potentially observable at the LHC by 
looking at the ratios of cross sections of the trilepton $+~\mET$ channels in certain
flavour specific modes. We show that even after considering possible leading standard 
model backgrounds these final states can lead to reasonable discovery significance at the
LHC with both $7$ TeV and $14$ TeV center-of-mass energy.}

\preprint{\hspace{6.5cm} FTUAM-12-80, IFT-UAM/CSIC-12-02, January 2012}

\date{\today}

\begin{document}

\section{Introduction}
Supersymmetry is one of the most well-motivated theories for explaining new physics beyond 
the standard model (SM) of particle physics. With the initiation of the large hadron 
collider (LHC) experiment at CERN, discovery of weak scale supersymmetric particles 
is highly envisaged. Moreover, the stable, neutral, and weakly interacting 
lightest supersymmetric particle (LSP) in an R-parity conserving theory can be an ideal
candidate for a non-baryonic dark matter for explaining the data from the WMAP 
satellite and large scale structure formation.

On the other hand, experimental evidences of neutrino oscillation have firmly established 
the necessity of massive neutrinos and the associated non-trivial mixing in the leptonic 
sector (see references \cite{Schwetz:2008er,GonzalezGarcia:2010er,Schwetz:2011qt} 
for the latest update on the three flavour global neutrino data). This is also a definite and 
very important indication of new physics because non-zero neutrino masses and mixing are
not included in the SM. 

Seesaw mechanism \cite{Minkowski:1977sc,Yanagida:1979as,Mohapatra:1979ia,Glashow:1979nm,
GellMann:1980vs,Schechter:1980gr,Schechter:1981cv} turns out to be the most simple
and natural \cite{Weinberg:1979sa,Weinberg:1980bf} way to accommodate tiny neutrino 
masses, $m_{\nu_i}$ ($\sum m_{\nu_i}<0.58$ eV \cite{Komatsu:2010fb}) both for supersymmetric 
and non-supersymmetric theories. However, neutrino mass models with canonical seesaw 
mechanism cannot be tested directly since the associated heavy states (to ensure smallness in 
neutrino masses) are usually $\sim$ ${\cal O} (10^{15}~{\rm GeV}$) and thus  well beyond the reach of 
any ongoing collider experiment. 

The inverse seesaw \cite{Mohapatra:1986aw,Mohapatra:1986bd,GonzalezGarcia:1988rw} 
is one of the viable alternatives with a seesaw mechanism operational at the TeV 
scale, which is well within the reach of the LHC. By proposal inverse seesaw mechanism relies on a 
new mass parameter, which breaks lepton number explicitly by two units, and at the same time small enough 
to ensure small neutrino masses without the requirement of any super heavy states. The smallness of 
this new lepton number breaking scale can arise either spontaneously \cite{GonzalezGarcia:1988rw}
or dynamically \cite{Bazzocchi:2009kc}. 

In the supersymmetric inverse seesaw model, $R$-parity \cite{Farrar:1978xj,Weinberg:1981wj,Sakai:1981pk,
Dimopoulos:1981dw} is conserved as a consequence of lepton number violation by two units. Thus the 
LSP is stable for this class of models and can be a viable candidate 
for the cold dark matter of the Universe.  

In this paper, we study signatures of the minimal supersymmetric version of the inverse seesaw model 
at the LHC.
This model can accommodate three flavour global neutrino data \cite{Schwetz:2008er,GonzalezGarcia:2010er} 
with a pair of SM gauge singlet superfields, which carry lepton number. These singlet scalars (having 
mixing with the doublet sneutrinos) can be thermal cold dark matter candidates, because the neutrino
Yukawa coupling in this model is $\sim$ ${\cal O}(10^{-1})$. In a 
supersymmetric inverse seesaw model just one pair of singlets can explain the observed neutrino
experimental data \cite{Hirsch:2009ra}. On the other hand, a non-supersymmetric inverse seesaw model 
requires at least two pairs of singlets \cite{Malinsky:2009df}. Apart from the neutrino 
masses and mixing, the supersymmetric inverse seesaw model has also been analysed earlier in the 
context of lepton flavour violation 
\cite{Deppisch:2004fa,Deppisch:2005zm,Ibanez:2009du,Hirsch:2009ra,Dev:2009aw}, 
leptogenesis \cite{Garayoa:2006xs,Gu:2010xc,Blanchet:2010kw}, dark matter 
\cite{Arina:2008bb,Kang:2011wb,An:2011uq} etc. 

In the case when one of the singlet sneutrinos is the LSP, the phenomenology at the LHC can be very
interesting. This is because in this model the neutrino Yukawa coupling is large and can lead to 
lepton flavor violating (LFV) coupling of the sneutrinos with a charged lepton and the chargino. These
couplings are related to the observed neutrino mixing angles and hence by studying the collider 
signatures of this model it is possible to study the relation between neutrino physics and the
physics at the high energy colliders.  

Supersymmetric particle searches from the 1 fb$^{-1}$ data, collected by ATLAS and CMS for $pp$ 
collision at center-of-mass energy, $\sqrt{s} = 7$ TeV, has found no significant 
signal over the expected SM background. In the
context of the constrained minimal supersymmetric standard model (CMSSM), 
searches by ATLAS exclude squarks and gluinos with masses below $950$ GeV \cite{Aad:2011ib}
at $95\%$ C.L. for some particular choice of other parameters. The results from CMS extend 
the mass limit to $1.1$ TeV \cite{Chatrchyan:2011zy,cms1,cms2}.
However, the third generation squarks can still be somewhat lighter, particularly in the context of a 
more general MSSM scenario. This is the reason, in this work, we choose to work with a spectrum where 
the squarks of the first two generations and the gluinos are very heavy ($\sim$ $1$ TeV) and the 
electroweak sector is relatively light so that the lighter chargino and neutralinos can be produced at 
the LHC. If in the production or in the decay chain the lighter chargino (${\tilde \chi}_1^\pm$) appears 
then it can have a decay into a charged lepton ($l$) via 
${\tilde \chi}_1^\pm \rightarrow l^\pm + {\tilde N}$, where ${\tilde N}$ represents the 
singlet sneutrino LSP. The ratios of the decay branching ratios into different charged lepton flavors
can be shown to correlate with the neutrino mixing angles \cite{Hirsch:2009ra}. 
Our aim in this paper is to look at these correlations by studying the trilepton + $\mET$ signature 
from the associated production of the lighter chargino (${\tilde \chi}_1^\pm$) and the second lightest 
neutralino (${\tilde \chi}^0_2$) at the LHC.

Similar correlations also appear in the decay of the LSP in the model of bilinear R-parity violation
\cite{Mukhopadhyaya:1998xj,Chun:1998ub,Choi:1999tq,Datta:1999xq,Romao:1999up,Porod:2000hv,Hirsch:2003fe,DeCampos:2010yu}, 
spontaneous R-parity violation \cite{Hirsch:2008ur} and in $\mu\nu$SSM \cite{Ghosh:2008yh,Bartl:2009an,Bandyopadhyay:2010cu}. 
In these models correlations with neutrino mixing angles have been studied in various context in the case of a
neutralino LSP decays as well as for other LSPs including the chargino \cite{Hirsch:2003fe}. The final states 
discussed in these cases generally include $multi-leptons ~+ ~jets ~+ ~\mET$ along with the presence of displaced 
vertices originating from the long-lived LSP. In the present paper we have studied this correlation in the 
decay of the NLSP chargino in minimal supersymmetric inverse seesaw model (MSISM) through the cleaner $trilepton ~+ ~\mET$ 
final state in the absence of any displaced vertex. 

It is known that at least two non-vanishing neutrino masses are 
essential \cite{Fukuda:1998mi,Ahmad:2002jz,Eguchi:2002dm} to account for oscillation data 
\cite{Schwetz:2008er,GonzalezGarcia:2010er}. In the minimal supersymmetric inverse seesaw 
model with one pair of singlets only one neutrino mass is generated at the 
tree level, while another neutrino mass is originating from the sneutrino-anti-sneutrino
loop \cite{Hirsch:1997vz,Grossman:1997is}. This feature is analogous to the $R$-parity 
violating (see review \cite{Barbier:2004ez} and references therein) models of light 
neutrino mass generation with bilinear terms \cite{Hall:1983id,Lee:1984kr,Lee:1984tn,
Hempfling:1995wj,Grossman:1997is,Grossman:1999hc,Hirsch:2000ef,Davidson:2000uc,Abada:2001zh,
Grossman:2003gq}. Issues of neutrino mass generation in MSISM have been addressed in 
ref. \cite{Hirsch:2009ra}. Similar issues have been discussed in refs. \cite{Deppisch:2004fa,
Arina:2008bb}, but with three generations of singlets. A model of neutrino mass generation
where the origin of neutrino mass is radiative and suppressed by inverse seesaw scale
has been advocated in ref. \cite{Ma:2009gu}.

In MSISM, the atmospheric neutrino mixing angle $(\theta_{23})$ correlates not only 
with the ratio of the branching ratios of the lighter chargino $(\chpm1)$ decay modes but also with ratio 
of the branching ratios of lepton flavour violating decays, $\tau/\mu \to \ell+ \gamma$ \cite{Hirsch:2009ra}.  
On the other hand, trilepton signals ($3l$, with or without tau lepton(s)) have been 
extensively studied for a long time as an important probe for supersymmetric 
models \cite{Frank:1995dj,Abbott:1997je,Barger:1998wn, Barger:1998hp,Matchev:1999nb,Baer:1999bq,
Matchev:1999yn,Bisset:2003ix,Sullivan:2008ki,Aaltonen:2008pv, Abazov:2009zi,Bhattacharyya:2009cc} 
(see also references [3,4,6,7] of ref. \cite{Barger:1998wn}). Besides, a hadronically quiet event like 
this always has the favour of reducible backgrounds. 
Moreover, multi-lepton signals have already been considered as an important probe
for seesaw models \cite{delAguila:2008cj,delAguila:2008hw,Chen:2011hc}.
Being motivated by these features together with the 
novel correlations mentioned earlier, we aim to perform a detailed analysis of 
trilepton ($3\ell,2\ell+1\tau$) + 
$\mET$~ signals for the MSISM taking into account possible SM backgrounds. 

As mentioned earlier, we search for the trilepton signatures, arising from the decay
of $\ntrl2 \chpm1$ pair. In our chosen parameter points the associated production of the lightest chargino 
with the next-to-lightest neutralino, $pp\to\ntrl2\chpm1 + X$ can occur with a detectable 
rate at the LHC. In addition, the lighter chargino and the second lightest neutralino
decays via two body leptonic modes with large branching ratios. The final state signal will produce three 
charged lepton and missing energy signature $(3\ell+\mET~$ or $2\ell+1\tau+\mET~)$, 
because of the presence of the stable singlet 
sneutrino LSP.  It is interesting to note that in our analysis the lightest neutralino $\ntrl1$ decays 
into a singlet scalar LSP and a light neutrino. Both of these decay products escape detection and 
thus $\ntrl1$ can be thought of as a virtual LSP, which also yields the missing energy signature at an 
accelerator experiment similar to that by an LSP. We investigate three body final states like 
$3\ell$ ($\ell=e,\mu$) and also $2\ell+\tau$-${\rm jet}$. Final states with more than one tau lepton have 
been dropped for small $\tau$ detection efficiency \cite{Bayatian:2006zz}. 
In the course of present 
analysis we choose to work with non-universal gaugino masses but maintain $M_2>M_1$,
where $M_1(M_2)$ are the soft masses for ${ U(1)}({ SU(2)})$ gaugino(s).
It has also been assumed that $\mu>M_2$, where $\mu$ 
is the coefficient of the only bilinear term in the superpotential of the minimal supersymmetric standard 
model (MSSM). With such a choice, $\ntrl2,\chpm1$ are essentially gaugino like. Moreover, 
since the 
first two generation squark masses are heavy, the process $pp\to\ntrl2\chpm1$ receives 
prime contributions from $W^\pm$-boson mediated processes. Three of our benchmark
points (BP1, BP3 and BP4 as defined later) are chosen with this criteria. We have, however, also considered 
the situation when $M_2>\mu$ for another benchmark point (BP2). However, in a situation
like this, $\ntrl2,\chpm1$ are higgsino like and consequently yield a smaller 
cross section for the process $pp\to\ntrl2\chpm1$. We show later that in a scenario
like this the trileptonic final state possesses lower significance compared to the $\mu>M_2$ scenario.
We will discuss this issue in more details later in section \ref{results}.
Having heavy squarks is also useful for 
suppressing flavour violating processes in the quark sector.

The ratio of the branching ratios for $\chpm1$ decaying into $\mu$ and $\tau$ channel in
MSISM shows sharp correlation with $\tan^2{\theta_{23}}$ \cite{Hirsch:2009ra}. 
It is clear that one of the three 
charged leptons appearing either in $3\ell+\mET$ or $2\ell+\tau+\mET$ final states must have 
its origin in $\chpm1$ decay. Using this idea we find that the ratio $\frac{\sigma(\mu^\pm+\sum\ell\ell)}
{\sigma(\tau^\pm+\sum\ell\ell)}$, with $\ell=e,\mu$ shows nice correlation with ${\rm tan}^2{\theta_{23}}$ 
even after the application of different kinematical cuts to reduce SM backgrounds. Definitely,
that $\mu$ and $\tau$ are coming from lightest chargino decay. Existence of this
final state correlation with neutrino mixing angle along with a large amount of $\mET$
provides a distinct signature for the MSISM.

The present paper is organized as follows. We start with a brief description of the 
underlying model in section \ref{model} and discuss the generation of small neutrino masses
as well as masses and mixing in the sneutrino sector. In section \ref{decays} we discuss the decay 
modes of the lighter chargino and the lighter neutralinos $(\ntrl1,\ntrl2)$ 
in our model and present a set of four benchmark points
(BPs) studied in this work. In addition we discuss how these decay modes can lead to the 
trilepton final states at the LHC. Section \ref{ev-gen} discusses
the details of the signal event generation and the background analysis. We present our results
in section \ref{results} and finally, our conclusions are provided in section \ref{conclusions}.


\section{Model}\label{model}
In the MSISM the particle content of the MSSM is extended by a pair of SM singlet fields, 
$\widehat\nu^c$ and $\widehat S$ having lepton numbers $-1$ and $+1$, respectively.
The model superpotential following refs. \cite{Arina:2008bb,Hirsch:2009ra} is written as
\beq
W= W_{MSSM}+\varepsilon_{ab}h^i_\nu\widehat{L}^a_i
\widehat{\nu}^c\widehat{H}_u^b+M_R\widehat{\nu}^c\widehat{S}+
\frac{1}{2}\mu_S\widehat{S}\widehat{S},
\label{Inverse-seesaw-superpot}
\eeq
where $W_{MSSM}$ is the MSSM superpotential. In eq. (\ref{Inverse-seesaw-superpot})
$\hat {L}_i$s are the $SU(2)_L$ doublet lepton superfields and $\hat H_u$ represents
a up-type Higgs superfield. $\hat \nu^c$ represents a right handed neutrino superfield
whereas $\hat S$ is another SM gauge-singlet superfield, but with non-zero lepton
number. In eq. (\ref{Inverse-seesaw-superpot}) coefficient of the lepton number 
violating term is given by $\mu_S$. In the limit $\mu_S$ $\to\ 0$, 
MSISM superpotential (see eq. (\ref{Inverse-seesaw-superpot})) restores lepton number conservation,
which is consistent with the t'Hooft naturalness criteria \cite{'tHooft:1979bh}.

The corresponding soft supersymmetry breaking Lagrangian is given by
\bea
-{\mathcal L}_{\rm soft} &=&-{\mathcal L}^{\rm MSSM}_{\rm soft} 
         +  m^2_{\nu^c} \widetilde\nu^{c\dagger} \widetilde\nu^c
         +m^2_S \widetilde S^{\dagger} \widetilde S \nn\\
&+& \left(\varepsilon_{ab} A^i_{h_\nu}
\widetilde L^a_i \widetilde\nu^c H_u^b +
B_{M_R} \widetilde\nu^c \widetilde S 
      +\frac{1}{2}B_{\mu_S}\widetilde S \widetilde S 
      +{\rm h.c.}\right),
\label{Inverse-seesaw-soft}
\eea
with $-{\mathcal L}^{\rm MSSM}_{\rm soft}$ representing soft terms of the MSSM.
Just like the coefficient $\mu_S$ appearing in the superpotential 
(eq. (\ref{Inverse-seesaw-superpot})), the $B_{\mu_S}$ parameter in the 
soft terms (eq. (\ref{Inverse-seesaw-soft})) violates lepton number
by two units. The MSISM thus includes two lepton number violating parameters.
Both of these will contribute to the Majorana neutrino mass matrix \cite{Hirsch:2009ra}.

Tree level neutrino mass matrix in the MSISM is a $5\times 5$ matrix in 
the basis $(\nu_l,\nu^c,S)$, where, $l\equiv e, \mu, \tau$. This is given by
\bea
\left( \begin{array}{ccccc} 
0 & 0 & 0 & m_{D_1} & 0 \\
0 & 0 & 0 & m_{D_2} & 0 \\
0 & 0 & 0 & m_{D_3} & 0 \\
m_{D_1} & m_{D_2} & m_{D_3} & 0 & M_R \\
0 & 0 & 0 & M_R & \mu_s 
\end{array}
\right),
\label{5by5numat}
\eea
where $m_{D_i} \equiv h^i_\nu v_u$ ($i = 1,2,3$) are the three light neutrino Dirac masses with $v_u$
as the vacuum expectation value of the up-type Higgs field, $\langle H^0_u \rangle$.
The quantities $h^i_\nu$ are the neutrino Yukawa couplings. The structure of the
matrix shown in eq. (\ref{5by5numat}) can be readily understood from 
eq. (\ref{Inverse-seesaw-superpot}) by looking at 
the mixing between different doublet and singlet neutral fermions. The effective 
3 $\times$ 3 mass matrix for the light neutrinos can be obtained as 
\beq
 (M^\nu_{\rm tree})_{ij}= \frac{\mu_S}{(M_R^2+\sum m^2_{D_k})}~m_{D_i}m_{D_j}.
\label{tree-neut-mass-matrix}
\eeq
In the seesaw approximation $(m_{D_i}<<M_R)$, the denominator of eq. (\ref{tree-neut-mass-matrix})  
becomes only $M^2_R$. The five mass eigenstates of eq. (\ref{5by5numat})
are denoted by $\widetilde n_i$, $i=1,\dots,5$, out of which $\widetilde n_{1,2,3}$
are nothing but three light neutrinos.

Structure of eq. (\ref{tree-neut-mass-matrix}) tells us that only one neutrino is massive at the
tree level with mass:
\beq
m_{\nu_3} = \frac{\mu_s}{(M_R^2+\sum m^2_{D_k})} \sum{m_{D_i}^2}.
\label{tree-level-neut-mass-matrix}
\eeq 
The smallness of the neutrino mass is ascribed to the smallness of the $\mu_s$ parameter, 
rather than the largeness of the Majorana-type mass, $M_R$, as required for the standard 
seesaw mechanism \cite{Nunokawa:2007qh}. With the choice of normal hierarchy in the light neutrino
masses this tree level mass will attribute to the atmospheric scale $\sim 10^{-11}$ GeV. 
In the regime of seesaw approximation, for a typical Dirac mass 
$m_{D_i}\sim 10^{2}$ GeV (assuming neutrino Yukawa couplings, 
$h_\nu^i \sim 10^{-1}$) and TeV scale $M_R$, the value of the parameter $\mu_s$ comes out to be 
$\sim 10^{-9}$ GeV. On the contrary, when $m_{D_i}\sim M_R$ (see eq. (\ref{tree-neut-mass-matrix}))
then the atmospheric neutrino scale ($\sim 10^{-11}$ GeV) is determined by $\mu_s$ only.

The neutrino mass matrix shown in eq. (\ref{tree-neut-mass-matrix}) is diagonalizable using a $3\times3$
unitary matrix as
\beq
 U^{{tr} T}~M^\nu_{tree}~U^{tr}= {diag}(0,0,m_{\nu_3}).
\label{tree-level-diag-neut}
\eeq
The matrix $U^{tr}$ contains information about the tree level neutrino mixing angles.
In order to satisfy the neutrino experimental data \cite{Schwetz:2008er,GonzalezGarcia:2010er,
Schwetz:2011qt} one requires a second non-zero neutrino
mass eigenvalue which can arise by including the one-loop corrections in the 
neutrino mass matrix \cite{Hirsch:2009ra}.

In this model the doublet and singlet sneutrinos mix after the electroweak symmetry breaking.
Thus the sneutrino mass squared matrix, $M^2_{\tilde\nu}$ is now a $10\times10$ matrix for 
MSISM, and assuming {\it CP} conservation this matrix can be decomposed into two 
$5\times5$ block matrices corresponding to {\it CP}-even and 
{\it CP}-odd sneutrino fields. The sneutrino mass term in the Lagrangian, then looks like,
\beq
{\mathcal L}_{\tilde\nu}=\frac{1}{2}~(\phi^R,\phi^I) \left(
\begin{array}{cc}
M^2_+ &  0 \\
0 &  M^2_- \\
\end{array} \right)\left(
\begin{array}{c}
\phi^R \\
\phi^I 
\end{array}\right),
\label{Lagrangian-neutral-scalar}
\eeq
where, $\phi^R=(\widetilde\nu^R_i,\widetilde\nu^{cR},\widetilde S^R)$,
$\phi^I=(\widetilde\nu^I_i,\widetilde\nu^{cI},\widetilde S^I)$.
The two mass squared matrices $M^2_\pm$ are given by \cite{Arina:2008bb}
\beq
M^2_\pm=
\left(\begin{array}{ccc}
(M^2_{\tilde L_{i}}+\frac{1}{2}M_Z^2 \cos2\beta+m^2_{D_i})\delta_{ij} & 
\pm(A^j_{h_\nu}v_u-\mu~ m_{D_j} \cot\beta) & m_{D_j} M_R\\[2mm]
\pm(A^i_{h_\nu}v_u-\mu~ m_{D_i} \cot\beta) & 
 m^2_{\nu^c}+M_R^2+\sum m^2_{D_k} &
 \mu_S M_R\pm B_{M_R}\\[2mm]
m_{D_i} M_R & \mu_S M_R\pm B_{M_R} &m_S^2+\mu^2_S+M_R^2\pm B_{\mu_S}
\end{array} \right),
\label{neutral-scalar-massmatrix}
\eeq
%
where $M^2_{\widetilde L_{i}}$ denote soft supersymmetry breaking mass squared terms for $SU(2)_L$ 
doublet sleptons and $M_Z$ is the Z-boson mass. The ratio of the two Higgs VEVs is defined as 
$\tan\beta = \frac{v_u}{v_d}$, where $v_d$ is the vacuum expectation value of the down-type Higgs 
field $H_d$. The real symmetric mass matrix of eq. (\ref{Lagrangian-neutral-scalar}) can be
diagonalized by a 10$\times$10 orthogonal matrix as follows
\beq
\bG ~ M^2_{\tilde\nu}~ \bG^T=
{\rm diag}(m^2_{\tilde N_1},\dots, m^2_{\tilde N_{10}}),
\label{scalar-mass-matrix-diag}
\eeq
with $m^2_{{\widetilde N}_1} < \dots < m^2_{{\widetilde N}_{10}}.$
Diagonalizing the {\it CP}-even and {\it CP}-odd mass matrices $M^2_\pm$ separately by 
\beq
\bGpm~ M^2_\pm~ \bGpm^T={\rm diag}(m^2_{{\widetilde N}_{i\pm}}),~i=1,\dots,5,
\label{CP-state-scalar-mass-matrix-diag}
\eeq
where ${\widetilde N}_{i+}$ and ${\widetilde N}_{i-}$ denote the $i$-th {\it CP}-even and {\it CP}-odd 
sneutrino mass eigenstates, respectively, leads to a different parameterization which can be used
in some cases. In this notation, for the set of chosen parameters 
(shown later) ${\widetilde N}_{1+}$  = ${\widetilde N}_2$ and
${\widetilde N}_{1-}$  = ${\widetilde N}_1$ and so on (see, eq. (\ref{scalar-mass-matrix-diag})).

       
\section{Decays of chargino and neutralino}\label{decays}
In this section we discuss the decays of charginos to charged leptons and singlet
sneutrinos as well as the decays of the lighter neutralinos. We shall also show how
these decays can lead to the final states, that we have proposed to study in this
paper. Our choices of the four benchmark points for a detailed collider study 
will also be presented here.
\subsection{Chargino decay}\label{Chargino-decay}
For the discussion of chargino decays we shall concentrate on a part of the 
parameter space where one of the singlet scalars of MSISM is the LSP. Hence this
scalar singlet will appear at the end of the supersymmetric cascade decay chains. 
For the present discussion let us assume that the dominant decay mode of the lighter 
chargino is in the two body mode
\bea
{\widetilde \chi}^\pm_1 \rightarrow {\widetilde N}_a + l^\pm_i, ~~~~~~a = 1,2, 
~~l_i = e, ~\mu, ~\tau,
\eea 
with ${\widetilde N}_1$ being the {\it CP} conjugated state to ${\widetilde N}_2$.  
The relevant piece of the Lagrangian for the calculation of this decay width is
\beq
{\mathcal\ L}_{\ell\widetilde\chi^-\widetilde\nu}=\overline{\widetilde\chi^+_j}(C^L_{ija}P_L 
+ C^R_{ija}P_R)l_i
\widetilde
N_a + h.c.~,
\label{chargino-decay-Lagrangian}
\eeq
where
\bea
C^L_{ija}&=&-\frac{1}{\sqrt{2}} [g\bV^*_{j1}(\bG_{ai} - i\bG_{a,i+5}) 
- h^i_\nu \bV^*_{j2}(\bG_{a4}
- i\bG_{a9})],\nn\\ 
C^R_{ija}&=& \frac{1}{\sqrt{2}} Y_{\ell_i} \bU_{j2}(\bG_{ai} - i\bG_{a,i+5}).
\label{chargino-couplings}
\eea
The $Y_{\ell_i}$s are the charged lepton Yukawa couplings and $\bU$, $\bV$ are two unitary 
$2\times2$ chargino mixing matrices such that $\bU^* m_{2\times2} \bV^{-1}=diag(m_{\chpm1},m_{\chpm2})$,
where $m_{\chpm1},m_{\chpm2}$ are the two physical chargino masses. The $2\times2$ mass matrix
$m_{2\times2}$ in the charged gaugino-higgsino basis $\psi^{+^T}=-i\widetilde{\lambda}^+_2,
\widetilde{H}^+_u,~\psi^{-^T}=-i\widetilde{\lambda}^-_2, \widetilde{H}^-_d$ is given by
\beq\label{chargino_mass_matrix}
m_{2\times2} =
\left(\begin{array}{ccccc}
M_2 & {g}{v_u} \\ \\
{g}{v_d} & {\mu} 
\end{array}\right).
\eeq
Here $g$ is the $SU(2)_L$ gauge coupling.

The corresponding decay widths are given as
\beq
\Gamma (\widetilde\chi^\pm_1\to\widetilde N_a + l^\pm_i)=\frac{(m^2_{\chi^\pm_1} - 
m^2_{\widetilde N_a})^2}{32\pi m^3_{\chi^\pm_1}} ({\mid C^L_{i1a}\mid}^2 + 
{\mid C^R_{i1a}\mid}^2).
\label{chargino-decay-width}
\eeq
The members of {\it CP} conjugated pair of sneutrinos being nearly mass 
degenerate ($m_{{\widetilde N}_1} \approx m_{{\widetilde N}_2}$) they are unlikely to be 
distinguished experimentally. Hence we sum over the {\it CP}-even and {\it CP}-odd sneutrino 
states of the {\it CP} conjugated pair. Thus
\beq
\Gamma (\widetilde\chi^\pm_1\to\widetilde N_{1+2} + l^\pm_i)\equiv 
\sum^2_{\alpha=1} \Gamma (\widetilde\chi^\pm_1\to\widetilde N_\alpha + l^\pm_i).
\label{decay-width-sum}
\eeq 
One can adjust the parameters $\mu_S$ and $B_{\mu_S}$ in such a way that the tree-level 
neutrino mass matrix contribution determines the atmospheric mass scale, while the one-loop 
corrections control the solar mass scale \cite{Hirsch:2009ra}. In such a situation it can be 
shown that in order to have small reactor neutrino mixing angle and maximal 
atmospheric neutrino mixing angle, the parameter $m_{D_1}$ has to be considerably smaller 
than other two Dirac masses and simultaneously, $m_{D_2} \sim m_{D_3}$.
The solar neutrino mixing angle can be kept large by keeping the parameters
$\delta_i \equiv A^i_{h_\nu} v_u - \mu m_{D_i} \cot\beta$ to be of the same order for all 
the three flavors, $i = e, ~\mu, ~\tau$. In this case, one can show that the decay width
of the lighter chargino,  
$\Gamma (\widetilde\chi^\pm_1\to\widetilde N_{1+2} + l^\pm_i)$ correlates with the corresponding
parameter $m^2_{D_i}$. The atmospheric neutrino mixing angle at the same time also behaves as 
$\tan^2 \theta_{23} \sim \frac{m^2_{D_2}}{m^2_{D_3}}$. Hence, one would expect that the ratio of the
branching ratios $\frac{Br(\widetilde\chi^\pm_1\to\widetilde N_{1+2}+\mu^\pm)}
{Br(\widetilde\chi^\pm_1\to\widetilde N_{1+2}+\tau^\pm)}$ must correlate with the ratio 
$\frac{m_{D_2}^2}{m_{D_3}^2}$. This has been shown in figure \ref{branching_frac_ratio}.
\FIGURE{\epsfig{file=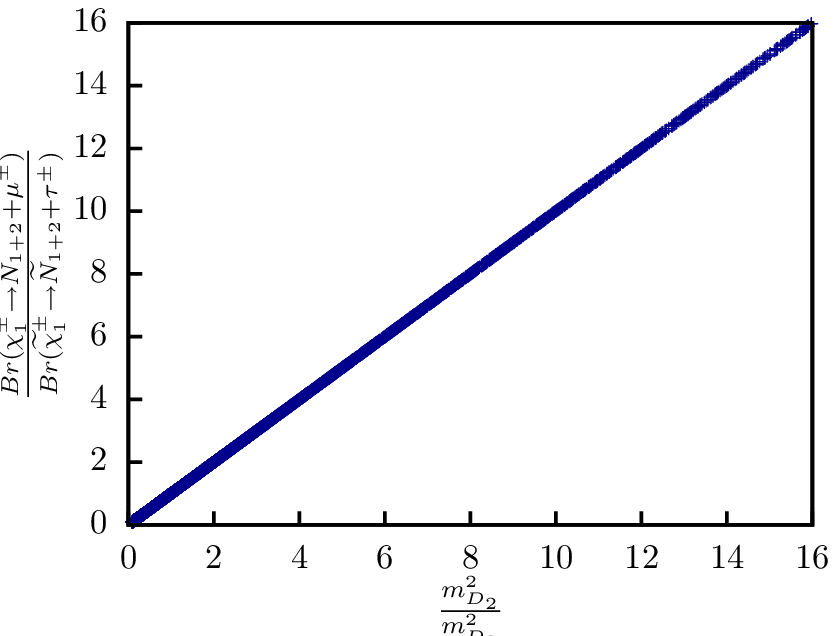,height=6.80cm} 
\caption{Correlation plot for the ratio of the branching ratios
 $\frac{Br(\widetilde\chi^\pm_1\to\widetilde N_{1+2}+\mu^\pm)}
{Br(\widetilde\chi^\pm_1\to\widetilde N_{1+2}+\tau^\pm)}$ with $\frac{m_{D_2}^2}{m_{D_3}^2}$.}
\label{branching_frac_ratio}}


\subsection{Neutralino decay}\label{neutralino-decay}
In our chosen benchmark points (defined later in this section) the lightest neutralino is 
the next-to-lightest supersymmetric particle (NLSP) and, 
decays dominantly through the two body decay channels 
$\widetilde\chi^0_1\to\nu_{l_i} +\widetilde N_{1,2},~l_i = e, \mu, \tau$.
The relevant interaction term of the Lagrangian is:
\beq
{\mathcal\ L}_{\nu\widetilde\chi^0\widetilde\nu}=\bar{\widetilde\chi}^0_j
(A^L_{mjb}P_L + A^R_{mjb}P_R)\nu_m \widetilde N_b + h.c.,
\label{neutralino-decay-Lagrangian}
\eeq
where
\bea
A^L_{mjb}&=&\frac{g}{2}(\bN^*_{j2} - \tan\theta_W \bN^*_{j1})
(\bG_{bi} - i\bG_{b(i+5)})U^{tr}_{im},\nn\\
A^R_{mjb}&=&-\frac{1}{\sqrt{2}}h^i_\nu U^{tr}_{im}\bN_{j4}(\bG_{b4} - i\bG_{b9}).
\label{neutralino coupling}
\eea
Here $g$ is the $SU(2)_L$ gauge coupling, $\theta_W $ is the weak mixing angle, and $\bN$
is the unitary 4x4 neutralino mixing matrix. Although the second lightest neutralino 
(${\widetilde \chi}^0_2$) decays mostly through the standard MSSM two-body charged lepton-slepton 
channel (${\widetilde \chi}^0_2 \to {\widetilde l_i}^\pm + l_i^\mp$), 
some of its branching fraction goes into the decay channels arising from the coupling given in 
eq. (\ref{neutralino-decay-Lagrangian}). Here, we have neglected the charged lepton flavor violating 
decay of ${\widetilde \chi}^0_2$. The decay width of a neutralino decaying into neutrino-sneutrino 
two-body mode is given as
\beq
\Gamma (\widetilde\chi^0_j\to\widetilde N_b + \nu_m)=\frac{(m^2_{\chi^0_j} - m^2_{\widetilde
N_b})^2}{32\pi m^3_{\chi^0_j}} ({\mid A^L_{mjb}\mid}^2 + {\mid A^R_{mjb}\mid}^2).
\label{neutralino-decay-width}
\eeq

\subsection{Trilepton signal and the benchmark points}\label{trilepton-benchmark}
In order to illustrate the trilepton signal we simulate $\widetilde\chi^0_2 
\widetilde\chi^\pm_1$ production followed by their two-body decays to produce 
$3\ell +~\mET$~ or~ $2\ell + \tau-{\rm jet} + \mET$~ final states, where $\ell = e, \mu$. 

\FIGURE{\epsfig{file=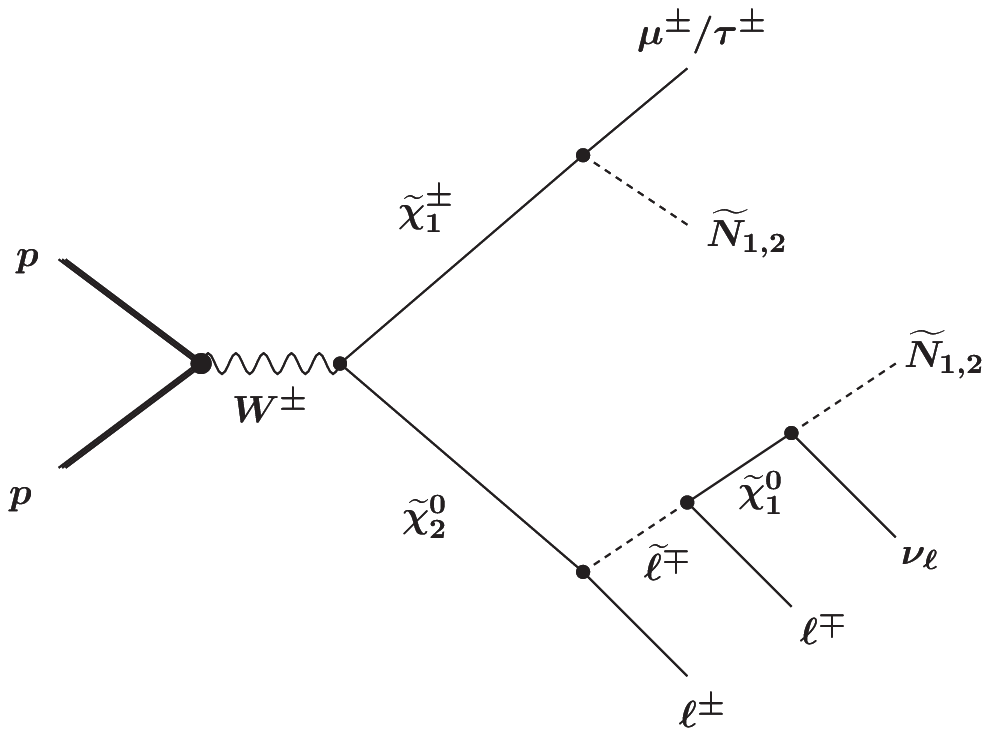,height=6.50cm} 
\caption{Feynman diagram for the process $pp\to 3\ell + \mET ~~{\rm{or}}~2\ell + \tau-{\rm{jet}}+ \mET$.}
\label{feyn-diag}}


As discussed above the production process and the decay cascades leading to these final states
are as follows
\bea
&&pp\to\widetilde\chi^0_2 + \widetilde\chi^\pm_1,\nn\\
&&\widetilde\chi^\pm_1\to\widetilde N_{1,2} +\mu^\pm/\tau^\pm,\nn\\ 
&&\widetilde\chi^0_2\to{\widetilde\ell^\pm} + \ell^\mp,\nn\\
&&\widetilde\ell^\pm\to\ell^\pm + \widetilde\chi^0_1,\nn\\
&&\widetilde\chi^0_1\to\nu_{l} +\widetilde N_{1,2}.
\label{signal-choice}
\eea
The Feynman diagram for the above mentioned final states is shown in figure \ref{feyn-diag}.
In the presence of heavy squarks ($\sim 1 ~TeV$), this is the leading
process for the chosen signal.

Because of the presence of the massive singlet sneutrino LSPs, $\widetilde N_{1,2}$, 
(quasi-degenerate in masses), we have, for this model, a large amount of 
missing energy in the final states. 
In order to have an appreciable signal rate one must have significant production cross section of 
${\widetilde \chi}^0_2 - {\widetilde \chi}_1^\pm$ pair and large branching ratios for the 
above-mentioned decays. To achieve these we have chosen four benchmark points (BPs) in the parameter space
where the detailed collider simulation has been performed. We scanned the whole parameter space 
to check for charged lepton flavour violating (LFV) decay widths and we found points both above and below
the experimental limits in different region of the parameter space. 
In all of the benchmark points, constraints from LFV decays \cite{Nakamura} are satisfied as well as
the atmospheric neutrino mixing is near maximal.
The input parameters for different benchmark points are given in table \ref{input1}.
The choices of the parameters $m^2_{D_i}$ will be shown later.
\TABLE{\begin{tabular}{|c c   c   c   c  c|} \hline 
 && BP1 & BP2 & BP3 & BP4 \\ \hline
tan$\beta$ &&5  &10 &20  &10      \\ 
$\mu$~(\rm{GeV}) &&330  &240 &280  &  350      \\
${M_1}$~(\rm{GeV}) &&170  &195  &160 &  240     \\ 
${M_2}$~(\rm{GeV}) &&220   &340 &240  &290   \\ 
${M_3}$~(\rm{GeV}) &&1100  &1100  &1100 &1100 \\
$M_R$~(\rm{GeV}) &&145  &160 &140  &150      \\ 
$\mu_s\times10^9$~(\rm{GeV}) &&$7.80$  &$7.81$ &$7.75$  &$7.76$      \\ 
$m^2_{\nu^c} ~\rm{(GeV^2)}$ &&2500   &3025   &2500   &3025  \\ 
$m^2_S\times 10^{-4} ~\rm{(GeV^2)}$ &&$4.0$    &$4.0$  &$4.0$  &$4.8$   \\ 
$B_{M_R} ~\rm{(GeV^2)}$ &&2500   &2500   &3500  &2500    \\ 
$B_{\mu_S} ~\rm{(GeV^2)}$ &&10  &10  &10  &10     \\
$M^2_{{\wt L_{i}}}\times 10^{-5}~\rm{(GeV^2)}$ &&$5.63$  &$5.63$  &$5.63$ &$5.63$     \\ 
$M^2_{{\wt e^c_{11}}}\times 10^{-4}~\rm{(GeV^2)}$ &&$2.99$    &$3.69$  &$2.59$  &$5.86$   \\ 
$M^2_{{\wt e^c_{22}}}\times 10^{-4}~\rm{(GeV^2)}$ &&$2.99$   &$3.69$    &$2.59$ &$5.86$   \\ 
$M^2_{{\wt e^c_{33}}}\times 10^{-4}~\rm{(GeV^2)}$ &&$3.53$   &$7.08$   &$7.90$  &$8.18$  \\ \hline
\end{tabular}
\caption{Values of the relevant input parameters for different benchmark points.
The quantities $M^2_{{\wt e^c_{ii}}}$ represent soft squared masses for the 
right-handed charged sleptons. Other parameters are defined in the text.\label{input1}}}

The mass splittings between the second lightest neutralino, the charged sleptons and the lightest
neutralino are maintained in a way, that the second lightest neutralino decays only through
charged lepton-slepton two body modes and the charged sleptons further decay into the lightest 
neutralino and charged lepton states. With these considerations we generated the sparticle spectrum using
{\tt{SuSpect}} (version 2.41)\cite{Djouadi:2002ze}. 
Masses of the neutrino and sneutrino states are computed 
using a self developed code in {\tt FORTRAN}. Relevant mass spectra for these benchmark
points are shown in table \ref{mass-spectrum}.

The choice of model parameters for different benchmark points are chosen to
yield statistically significant final states. As an illustrative
example, production cross sections for the $\ntrl2\chpm1$ pair with $7$ TeV
center of mass energy at LHC are in the range of $200-300$ fb for the first
and third benchmark points. For the fourth benchmark point with relatively heavy 
$\ntrl2\chpm1$ pair (see table \ref{mass-spectrum}) the production
cross section is reduced by a factor of $4(3)$ compared to
the first(third) benchmark point. On the contrary, a higgsino like
$\ntrl2\chpm1$ pair (BP2) yield a similar production cross-section
like BP4, in spite of having a lighter $\ntrl2\chpm1$ pair.
Thus, the region of parameter space with higgsino like
$\ntrl2\chpm1$ pair is unfavorable for this analysis.

\TABLE {\begin{tabular}{| c  c  c  c c |} \hline 
  & BP1 & BP2 & BP3  & BP4\\ \hline 
 &  &  &  & \\
$\widetilde N_{1+}$ &153.27  &169.18  &147.96  &159.76 \\
$\widetilde N_{2+}$ &247.38  &256.45  &244.81 &266.53  \\
$\widetilde N_{1-}$ &153.27  &169.18  &147.96 &159.76  \\
$\widetilde N_{2-}$ &247.34  &256.42  &244.76  &266.50 \\
${\wt n}_4$ &145.46  &160.52 &140.42  &150.45 \\ 
${\wt n}_5$ &145.46  &160.52  &140.42 &150.45\\
$\selL,\smuL$ &751.28 &751.37  &751.39  &751.37  \\ 
$\selR,\smuR$ &178.00  &196.97  &166.83  &245.98 \\ 
$\stau1$ &193.11  &269.48  &284.12  &289.17  \\
$\stau2$ &751.30  &751.40  &751.53  &751.43  \\ 
$\ntrl1$ &159.51  &172.90  &151.80  &226.36 \\
$\ntrl2$ &198.11  &234.59  &207.67  &264.16\\
${\chpm1}$ &192.79  &215.47  &203.72 &255.99\\
${\chpm2}$ &363.91  &372.89  &326.77 &391.60\\ \hline
\end{tabular}
\caption{Relevant mass spectra obtained for four benchmark points
with ${\wt n}_{1,2,3}=\nu_{1,2,3}$.\label{mass-spectrum}}}
%
Note that hadronically quiet trilepton signal ($3 \ell + \mET$~) will get very little contribution 
from squark-squark, squark-gluino and gluino-gluino pair production. On the other hand, when we have 
$2\ell + \tau$-jet + $\mET~$ signal, then one should consider all other sources of dilepton 
+ 1-jet + $\mET$ events where one jet can be faked as a $\tau$-jet. For example, one can have a
jet out of a squark decay (${\tilde q} \to q^\prime + {\widetilde \chi}_1^\pm$) from one side of the 
cascade. However, since in this model the squarks are much heavier ($\sim 1~\rm{TeV}$) and after 
incorporating the probability of any jet faking as a $\tau$-jet, the event rate comes out to be 
negligibly small compared to the one generated from chargino-neutralino production. Hence the main 
contribution to $2\ell+\tau$-jet + $\mET$~ signal comes from ${\widetilde \chi}^0_2{\widetilde \chi}^\pm_1$ production only. 


\section{Event generation and \\
background analysis}\label{ev-gen}
On the basis of the discussion presented in the previous section, let us now provide a 
detailed description of event generation and subsequently, the background analysis.
The decay widths corresponding to the two-body modes shown in eq. (\ref{signal-choice}) have been 
used to modify the branching fractions of the charginos and neutralinos obtained from 
{\tt{SuSpect}}. These input files are then fed to {\tt{PYTHIA}} (version 6.409) \cite{Pythia6.4} 
for event generation and showering. Initial and final state radiation, decay, hadronization, 
fragmentation and jet formation are implemented following the standard procedures 
in {\tt{PYTHIA}}. Factorization and renormalization scales are set at $\sqrt{\widehat s}~$
(i.e $\mu_R =\mu_F =\sqrt{\widehat s}$~), where $\sqrt{\widehat s}~$ is the parton level centre of mass energy. 
We have used the leading order CTEQ5L parton distribution 
functions \cite{Lai:1999wy,Pumplin:2002vw} for the colliding protons. Some of the background 
events are generated using {\tt{ALPGEN}} (version 2.14) \cite{Mangano:2002ea} with default 
factorization and renormalization scales. The jets are constructed using cone algorithm in PYCELL. 
Only those jets are constructed which have $p_T > 20~{\rm GeV}$ and $\mid\eta\mid < 2.5$. 
To simulate detector effects we have taken into account smearing of jet energies by a Gaussian 
probability density function of width \cite{Barr:2009wu} 
$\sigma (E)/E_j = (0.6/\sqrt{E_j[GeV]}) + 0.03$ where $E_j$ is the unsmeared jet energy.  

In order to find three isolated leptons in the final states we impose following cuts and  
isolation criteria:\\
\hspace*{0.5cm}\vspace*{0.2cm} I.~Leptonic events  are selected only if $p^\ell_T>8 ~{\rm{GeV}}$ and $\mid\eta^\ell\mid<2.4$.\\
\hspace*{0.5cm}\vspace*{0.2cm} II.~Lepton-lepton separation $\Delta R(\ell, \ell)$ set to be $> 0.2$, where $\Delta R=
\sqrt{(\Delta\eta)^2 + (\Delta\phi)^2}$.\\
\hspace*{0.5cm}\vspace*{0.2cm} III.~Lepton-jet separation $\Delta R(\ell, j)$ chosen to be $> 0.5$.\\
\hspace*{0.5cm} IV.~The sum of $E_T$ deposits of the hadrons which fall within a cone of $\Delta R\le 0.2$ around 
\hspace*{0.5cm}\vspace*{0.2cm} a lepton, must be less than $10~ {\rm GeV}$.

A $p_T$ cut of $10~{\rm GeV}$ and $17~{\rm GeV}$ \cite{Bayatian:2006zz} is applied on final state muons 
and electrons respectively, for the analysis at $7~{\rm TeV}$ and $14~{\rm TeV}$ center of mass energies 
at the LHC. The $\tau$-jets are counted with $p_T\ge 20 ~{\rm GeV}$ and $\mid\eta^\tau\mid<2.4$. 
The $\tau$'s are then counted according to the visible energy bins. A $\tau$-jet is treated as 
tagged or untagged according to the efficiency ($\epsilon_\tau$) of the most efficient algorithm 
given in \cite{Tau-eff}. In reference \cite{Tau-eff}, $\tau$ identification efficiency 
obtained from actual collision data at $7~{\rm TeV}$ center of mass energy has also been quoted.
The efficiencies obtained from Monte-Carlo simulation and from the data agrees very well. However, 
for higher luminosity with $14~{\rm TeV}$ center of mass energy, a lot of underlying 
events are expected to be there, which can perhaps bring down the detection efficiency. In this case also we have used the same efficiency as in $7~{\rm TeV}$ case hoping the experimentalists can maintain the efficiency as we have now. Unlike $\tau$, detection efficiencies of $e$ and $\mu$ are assumed to be  $100\%$. 

We have analysed the SM backgrounds in some detail. The dominant background events arise from $t\bar t$ 
and $WZ$ production at the LHC. Apart from these, contributions from $ZZ$, $WW$, $Zb\bar b$, $Wb\bar b$, 
$Z+jets$, $Wt$, $tb$, $WWW$, $Wt\bar t$ events have also been studied at the leading order. We also 
studied $QCD$ di-jet events. But after putting the cuts to reduce backgrounds as mentioned below we 
found no trilepton events for $1~{\rm fb^{-1}}$ integrated luminosity from these particular $QCD$ 
events. We use {\tt{ALPGEN}} for an estimation of $Zb\bar b$, $Wb\bar b$, $Wt$, $tb$, $Z+jets$, $WWW$, 
$Wt\bar t$ backgrounds. We generate these events at the parton level using {\tt{ALPGEN}} and fed 
those partonic events to {\tt{PYTHIA}} for showering, hadronization, fragmentation, decay, etc. 
The other events are generated and analysed using {\tt{PYTHIA}}. It should be mentioned that the 
importance of these processes have already been emphasized in the literature\cite{Sullivan:2008ki,Bhattacharyya:2008zi}. 

The trilepton signal in our model arising out of chargino-neutralino production is accompanied by 
large missing transverse energy ($\mET$~), because of a pair of singlet sneutrino LSPs and a neutrino. 
As an example, the $\mET~$ spectrum of background events as well as the signal events ($3\mu + \mET$~) 
for the first benchmark point (BP1) are shown in figure \ref{missing_pt}.
\FIGURE{\epsfig{file=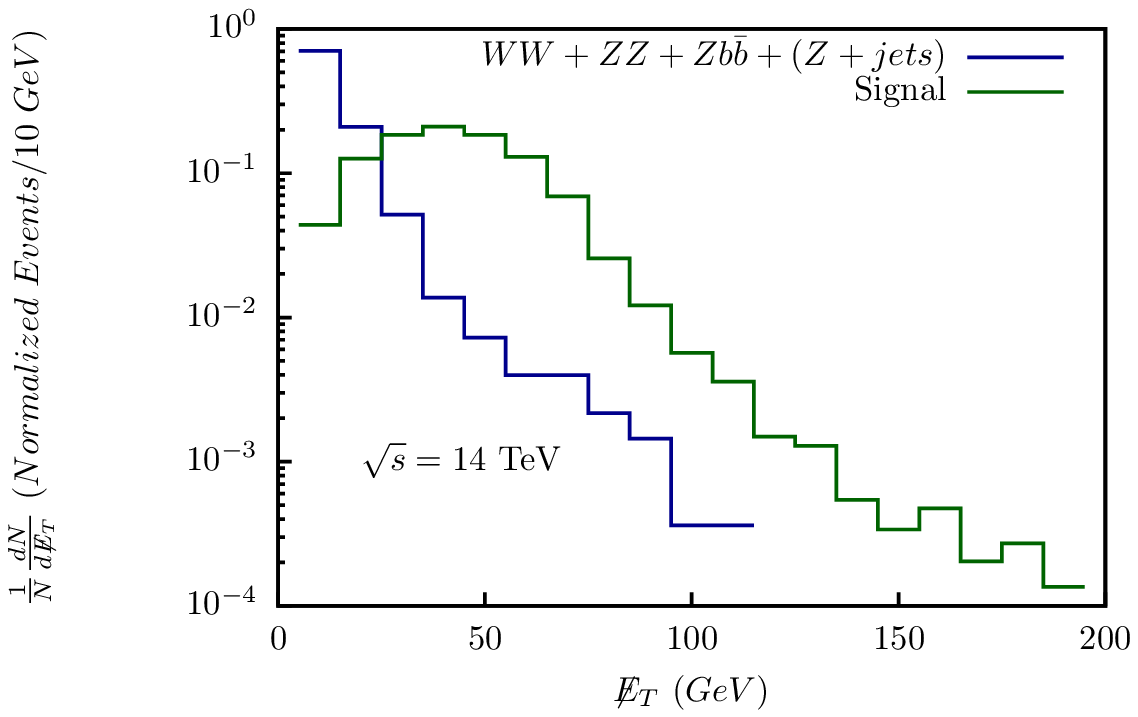,height=7.15cm} 
\caption{$\mET$ plot for signal events $3 \mu + \mET$  and summed up 
contribution coming from the $WW, ZZ, Zb\bar b, Z+1~jet, Z+2~jets$ background for LHC at 
$14~{\rm TeV}$ center of mass energy with $1$ fb$^{-1}$ of integrated luminosity. The 
$\mET$~ bin size is chosen to be $10$ GeV.}
\label{missing_pt}}

These distributions are obtained without applying any cuts to reduce background events. 
It is evident from the plot in figure \ref{missing_pt} that a strong $\mET$ cut will 
affect the signal cross-section very mildly, but it reduces significantly background events
coming from some processes. Therefore, a cut $\mET > 25~{\rm GeV}$ is applied for  
background rejection. For some other channels; $t\bar t, WZ, Wt\bar t, WWW$ the $\mET$~
distributions do not peak before $25~ {\rm GeV}$ as shown in figure \ref{missing_pt1}.
\FIGURE{\epsfig{file=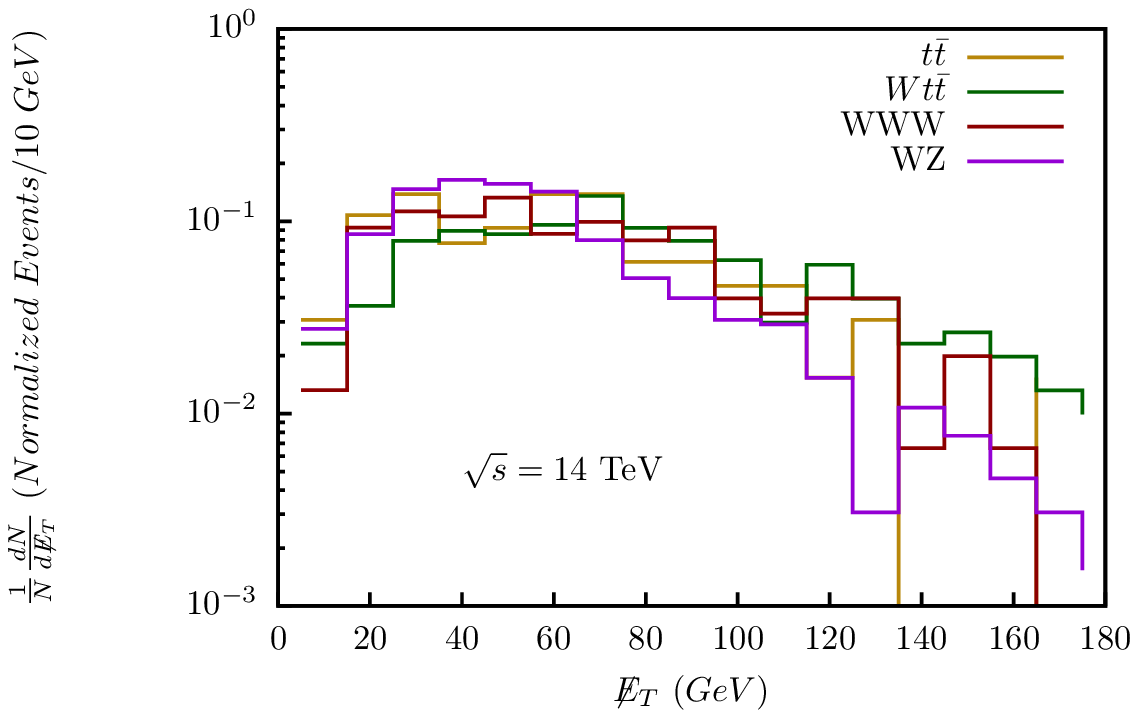,height=7.15cm} 
\caption{$\mET~$ plot for $3 \mu + \mET~$ events obtained from $t\bar t, WZ, Wt\bar t, WWW$ backgrounds 
for LHC at $14~{\rm TeV}$ center of mass energy with $1$ fb$^{-1}$ of integrated luminosity. 
The $\mET$~ bin size is chosen to be $10$ GeV.}
\label{missing_pt1}}
Hence, the above mentioned $\mET~$ cut does not seriously affect these background events. To 
reduce these events we have further applied two more cuts. An invariant mass cut on the 
opposite sign dilepton pair, $80~{\rm GeV} > M^{\ell\ell}_{inv} > 100~{\rm GeV}$ removes 
backgrounds coming from $Z$-bosons. To manifest this idea we show invariant mass distribution 
in figure \ref {inv_mass} constructed from opposite sign muon pairs for signal events and $WZ$ 
background events.

\FIGURE{\epsfig{file=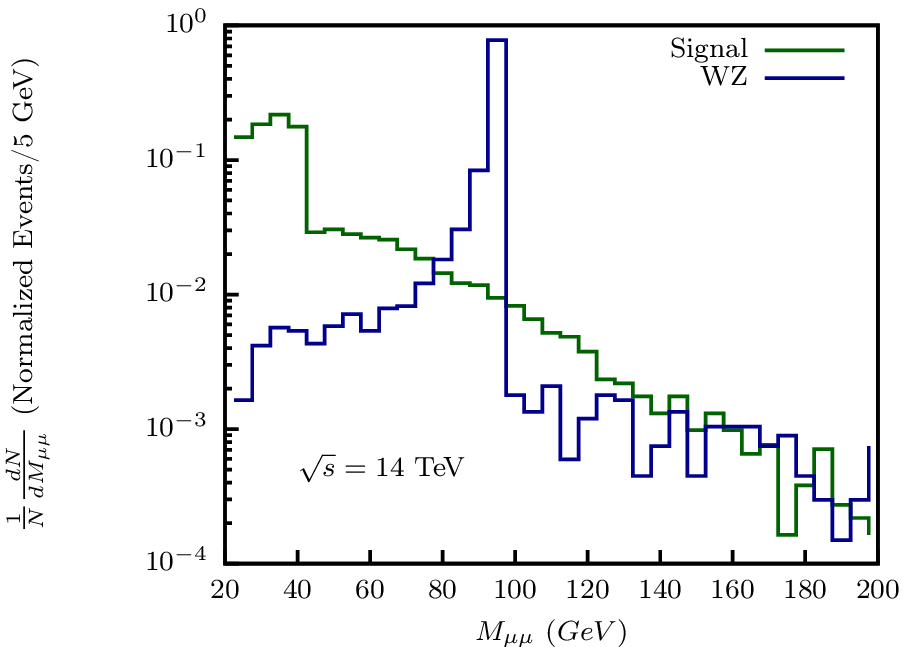,height=6.80cm} 
\caption{$M_{\mu\mu}~$ plot for signal events and $WZ$ background 
for LHC at $14~{\rm TeV}$ center of mass energy. The $M_{\mu\mu}~$ bin size is chosen to be $5$ GeV.}
\label{inv_mass}}
On the other hand, rejection of tagged $b$-jet events significantly reduces 
backgrounds coming from $t\bar t$ events. A jet (with $\mid\eta\mid<2.5$) is reconstructed 
as a $b$-jet if the $\Delta R$ separation between the jet and the $b$-quark (with $p_T > 5~{\rm GeV}$) 
is less than $0.2$. The b-jet identification efficiency is taken to be $50 \%$.
 
In order to perform the collider analysis we have randomly generated $m^2_{D_i}$ and $\delta^2_i$ within 
certain range: $(\sum_i m^2_{D_i})^{1/2} \in 10^{[-4,2.6]}$ and 
$(\sum_i \delta^2_i)^{1/4} \in 10^{[-4,3]}$ \cite{Hirsch:2009ra}.
Moreover, we also consider $(\sum m^2_{D_i})^{1/2}>(\sum_i \delta^2_i)^{1/4}$,
such that Dirac neutrino masses give the dominant contribution to
the chargino decay \cite{Hirsch:2009ra}.
Around each of the four benchmark points we select a set of six to seven
points of these randomly generated parameters. These points will be useful
for the correlation study discussed later in section \ref{results}.
Remember that these parameters control the neutrino masses and the mixing angles and 
our choices of benchmark points are such that the atmospheric neutrino mass scale is 
determined by the tree level neutrino mass matrix contribution. Before showering in 
{\tt{PYTHIA}}, as mentioned earlier, the ratio 
$\frac{Br(\widetilde\chi^\pm_1\to \widetilde N_{1+2} + \mu^\pm)}{Br(\widetilde\chi^\pm_1\to 
\widetilde N_{1+2} +\tau^\pm)}$ shows a very nice sharp correlation when plotted against 
$\frac{m^2_{D_2}}{m^2_{D_3}}$ which is a measure of $\tan^2 \theta_{23}$. We have done the showering 
for four benchmark points introduced in table \ref{input1} and 
table \ref{mass-spectrum} to look for the ratio 
$\frac{\sigma(pp \to \mu \sum \ell \ell + \mET~)}{\sigma(pp \to \tau \sum \ell \ell + \mET~)}$ 
with $\ell=e,\mu$. Since one $\mu$ and one $\tau$ in these final states always come 
from the decay of ${\widetilde \chi}_1^\pm$, we would expect that this ratio will also go
as $\sim \tan^2\theta_{\rm 23}$. Hence, by measuring this ratio from the trilepton signals 
one can obtain information about the atmospheric neutrino mixing angle at the LHC. 
On the other hand, a precise measurement of the atmospheric neutrino 
mixing angle at the oscillation experiments can be used to predict the allowed range of the above 
ratio at the LHC. In the following section we give a quantitative estimate of this ratio
for our choices of benchmark points (along with randomly selected values of $m^2_{D_i}$) 
and show that for each of these points the various signal events included in the calculation
of this ratio can be statistically significant. 
\section{Results}\label{results}
In order to study the correlation between the atmospheric neutrino mixing angle ($\theta_{23}$)
and the final states with trilepton + $\mET$ at the LHC, we look at the ratio of 
cross sections $\frac{\sigma(pp \to \mu \sum \ell \ell + \mET~)}
{\sigma(pp \to \tau \sum \ell \ell + \mET~)}$,
$\ell = e,\mu$. As mentioned in the introduction, in the denominator the $\tau$ must always come from 
the decay of ${\tilde \chi}^\pm_1$ because we are considering final states with only one $\tau$-jet and 
neglecting lepton flavor violating decays of ${\tilde \chi}^0_2$ and ${\tilde \ell}^\pm$. For the same reason,
in the numerator one $\mu$ must always also come from the decay of ${\tilde \chi}^\pm_1$. Hence, naively we 
would expect that this ratio of cross sections will also show nice correlation with the atmospheric neutrino 
mixing angle $\theta_{23}$. 

After applying different cuts to reduce backgrounds and taking into 
account the $\tau$-tagging efficiency, we find that the ratio of trilepton signal cross section
again shows a nice correlation with the atmospheric neutrino mixing angle $\tan^2\theta_{\rm 23}$. 
However, in this case the numbers change from the ratio of branching ratios, discussed earlier and 
the straight lines obtained are steeper than the one shown in figure \ref{branching_frac_ratio}. This 
happens because in our simulation we take the detection efficiency of $\mu$ to be $100 \%$ as opposed to
the $\tau$ detection efficiency, which is smaller \cite{Bayatian:2006zz}. Since the branching fractions 
of $\tau$ events are in the denominator of the ratio, the numbers naturally go up. 

The cross-sections and the corresponding statistical significance ($\frac{S_x}{\sqrt{B_x+S_x}}$
with $x=e,\mu,\tau$) obtained from our simulation for LHC are shown in this section. 
Here $S_x$ is defined as the number of $x\sum\ell\ell$ signal events and $B_x$ 
is defined as the number of corresponding background events. 
In more simple form significance for the $\mu\sum\ell\ell+\mET~$ channel
is defined as $\frac{S_{\mu e e} + S_{\mu \mu\mu}}{\sqrt{S_{\mu e e} + S_{\mu \mu\mu}
+B_{\mu e e} + B_{\mu \mu\mu}}}$. In a similar fashion significance 
for the $\tau\sum\ell\ell+\mET~$ channel can be obtained.

We quote the results below for an integrated luminosity of 25 ${\rm fb}^{-1}$ for the LHC with 
$7~{\rm TeV}$ and $14~{\rm TeV}$ center-of-mass energies. The results are obtained with the cuts 
mentioned in section \ref{ev-gen}. Throughout this analysis we have used leading order cross sections 
for the signals as well as all the backgrounds at the LHC. However, if next-to-leading order (NLO) corrections 
are included the statistical significance will not change much. For example, if NLO corrections 
are included the signal cross section at 14 TeV LHC is expected to increase by 1.25 to 1.35 \cite{Beenakker:1999xh}. 
As discussed above, a large contribution to the background comes from the $t {\bar t}$ events. The NLO cross section 
for $t {\bar t}$ production at 14 TeV LHC is about 800 pb \cite{Kidonakis:2008mu,Cacciari:2003fi} which is about a 
factor of two larger than the leading order cross section that we have used in our analysis. Thus taking into account 
the NLO contribution of all the major background events along with the signal event, the significance 
$S_x/\sqrt{B_x+S_x}$ estimated for our signal, will not change much and remains conservative in 
comparison to the uncertainties in the production cross sections. 

Values of the randomly generated parameters $m^2_{D_i}$,
for four chosen benchmark points are presented in
table \ref{input2}. For the numerical analysis we choose to vary
$m^2_{D_1}$ in the range of $10^{-4}-10^{-2}$ GeV$^2$, whereas $m^2_{D_{2,3}}$
are varied within $10^{-2}$ to $10^{2}$ GeV$^2$. The $\delta^2_i$ are also
varied accordingly, but keeping the constraints $(\sum m^2_{D_i})^{1/2}
>(\sum \delta^2_{i})^{1/4}$. The scale of $m^2_{D_i}$ has a strong influence on 
the decay processes $\chpm1\to\widetilde N_{1+2}+\mu^\pm/\tau^\pm$ and
$\widetilde\chi^0_j\to \widetilde N_b+\nu_m$. In order to achieve a statistically 
significant trilepton final state originating from $\ntrl2\chpm1$ pair,
we would like to have $Br(\chpm1\to\widetilde N_{1+2}+\mu^\pm/\tau^\pm)$
to be large and $Br(\ntrl2\to \widetilde N_b+\nu_m)$ to be small, simultaneously.
However, in the limit $m_{D_i}\sim M_R\sim$ $\cal{O}$ $(10^2~\rm{GeV})$, the neutrino Yukawa couplings 
$h^i_\nu$ are $\sim$ $\cal{O}$ $(1)$. Then as can be seen from eqs. (\ref{chargino-decay-width})
and (\ref{neutralino-decay-width}) both of these decay widths are large and
consequently, yields a smaller  branching ratio for $\ntrl2\to \widetilde{\ell}^\pm+\ell^\mp$.
We observe that in this case it is rather difficult to achieve a statistically significant
final state particularly for the $\tau\sum\ell\ell+\mET~$ mode.

\TABLE {\begin{tabular}{|c c c c c|} \hline 
$m^2_{D_i}~\rm{(GeV^2)}$ & BP1 & BP2 & BP3 & BP4\\ \hline 
$m^2_{D_1}\times10^4$ & $2.12$ & $3.30$ & $1.58$ & $4.24$  \\
$m^2_{D_2}$ & 71.80 & 62.96 & 68.50 & 80.33 \\
$m^2_{D_3}$ & 62.45 & 54.87 & 66.63 & 86.00\\ \hline
\end{tabular}
\caption{Randomly generated values of $m^2_{D_i}$ 
corresponding to the four benchmark points as indicated in section \ref{decays}. 
\label{input2}}}
%

In the trilepton signals studied in this work, one lepton comes from the lighter chargino 
(${\widetilde \chi}^\pm_1$) decay and the other two same flavour opposite sign leptons 
come from the second lightest neutralino (${\widetilde \chi}^0_2$) decay. Since the 
probability of getting electrons from the chargino decay is suppressed compared to muons 
or taus, events with odd number of electrons ($e e e$ and $e\mu\mu$) should have smaller 
cross-sections compared to others, which is clearly reflected in the signal cross-sections. 
This feature is intrinsically related with the small but non-zero reactor neutrino 
angle \cite{Abe:2011sj}, which will be discussed again later.
In table \ref{cross-section-lhc7} and \ref{cross-section-lhc14} chosen trilepton $+\mET~$ cross
sections are shown along with the total standard model background cross section for the
LHC at center-of-mass energy, $\sqrt{s}=7$ and $14$ TeV, respectively. The corresponding
statistical significance of the signals are shown respectively in table \ref{significance-lhc7}
and table \ref{significance-lhc14}. We can see 
from table \ref{significance-lhc7} that, at the LHC even with $\sqrt{s}=7$ TeV, the lowest signal
significance for $\tau\sum\ell\ell+\mET~$ final state, that we have obtained, 
is greater than $3\sigma$ for an integrated luminosity of $25$ fb$^{-1}$. Hence, the trilepton
$+\mET~$ data for $25$ fb$^{-1}$ integrated luminosity at $7$ TeV LHC should be able to constrain the
theoretical parameter space of this model. These numbers (significance) are 
much higher for LHC with $\sqrt{s}=14$ TeV and are shown in table \ref{significance-lhc14}.
It is once again evident from these tables that a higgsino like $\ntrl2\chpm1$ pair (BP2)
yields statistically less significant specific trilepton final state. 
In other words for such benchmark points, the significance of the 
final state trilepton signal is less promising.
This situation is comparable to a heavy gaugino like
$\ntrl2\chpm1$ pair as represented by BP4.

\TABLE {\begin{tabular}{|ccccccc|} \hline 
Tri-lepton & & signal& & & &Background\\ 
events & & $\sigma$ (fb) & & & &$\sigma$ (fb) \\ \hline 
&BP1&BP2&BP3&BP4 & &\\ \hline
$e e e$      &0.37  &0.31  &0.50 &0.23 & &8.73  \\
$e e \mu$    &9.47  &5.37  &8.30 &3.12 & &18.91  \\
$e\mu\mu$    &1.08  &0.49  &1.26 &0.66 & &21.15  \\
$\mu\mu\mu$  &24.13  &8.21  &18.85 &8.51& &23.84  \\
$e e \tau$   &2.93  &2.14  &2.86 &1.40 & &4.60  \\
$\mu\mu\tau$ &7.17  &3.39  &6.99 &4.04 & &13.18\\ \hline
\end{tabular}
\caption{Cross-section for different trilepton channels are shown here 
for four different benchmark points along with 
their total SM background contribution for LHC with $\sqrt{s}=7$ TeV. 
Corresponding input parameters and mass spectrum are given in
table \ref{input1}, table \ref{input2} and table \ref{mass-spectrum}, respectively.
.\label{cross-section-lhc7}}}
\TABLE {\begin{tabular}{|c c c c c c|} \hline 
&Tri-lepton & & Significance && \\ 
&events & & $\frac{S_{x\sum\ell\ell}}{\sqrt{B_{x\sum\ell\ell}+S_{x\sum\ell\ell}}}$ && \\ \hline 
&&BP1&BP2&BP3&BP4 \\ \hline
&$\mu e e + \mu \mu \mu$        &19.23 &9.05 &16.24 &7.89\\
&$\tau e e + \tau \mu \mu$      &9.57 &5.72 &9.37 &5.65 \\ \hline
\end{tabular}
\caption{Statistical significance of the studied trilepton signals with integrated luminosity 
$25~{\rm fb^{-1}}$ at the LHC for $\sqrt{s}=7$ TeV for different benchmark points. 
\label{significance-lhc7}}}

\TABLE {\begin{tabular}{|ccccccc|} \hline 
Tri-lepton & & signal& & & &Background\\ 
events & & $\sigma$ (fb) & & & &$\sigma$ (fb) \\ \hline 
&BP1&BP2&BP3&BP4 & &\\ \hline
$e e e$       &1.29  &0.91  &1.56 &0.75  & &24.92  \\
$e e \mu$     &27.63  &15.65  &23.50 &10.48 & &91.64  \\
$e\mu\mu$     &3.00  &1.66  &3.59 &2.14 & &117.97  \\
$\mu\mu\mu$   &65.42  &24.32  &53.10 &26.47 & &85.94  \\
$e e \tau$    &8.82  &6.48  &9.11 &4.76 & &29.16  \\
$\mu\mu\tau$  &20.11  &9.80  &18.46 &12.48 & &56.29\\ \hline
\end{tabular}
\caption{Cross-section for different trilepton channels are shown here 
for four different benchmark points along with 
their total SM background contribution for LHC with $\sqrt{s}=14$ TeV. 
Corresponding input parameters and mass spectrum are given in
table \ref{input1}, table \ref{input2} and table \ref{mass-spectrum}, respectively. 
\label{cross-section-lhc14}}}
\TABLE {\begin{tabular}{|c c c c c|} \hline 
Tri-lepton & & Significance && \\ 
events & & $\frac{S_{x\sum\ell\ell}}{\sqrt{B_{x\sum\ell\ell}+S_{x\sum\ell\ell}}}$ && \\ \hline 
&BP1&BP2&BP3&BP4 \\ \hline
$\mu e e + \mu \mu \mu$       &28.28 &13.55 &24.02 &12.61 \\
$\tau e e + \tau \mu \mu$       &13.53 &8.07 &12.97 & 8.50\\ \hline
\end{tabular}
\caption{Statistical significance of the studied trilepton signals with integrated luminosity 
$25~{\rm fb^{-1}}$ at the LHC for $\sqrt{s}=14$ TeV for different benchmark points. 
\label{significance-lhc14}}}

We present the correlation plots, obtained with different randomly generated
values of $m^2_{D_i}$ and $\delta^2_i$ around each of the four benchmark points.
These are shown in figure \ref{correlation_plot1} and figure \ref{correlation_plot2}
for the LHC with $\sqrt{s}=7$ TeV and $14$ TeV, respectively.
\FIGURE{\epsfig{file=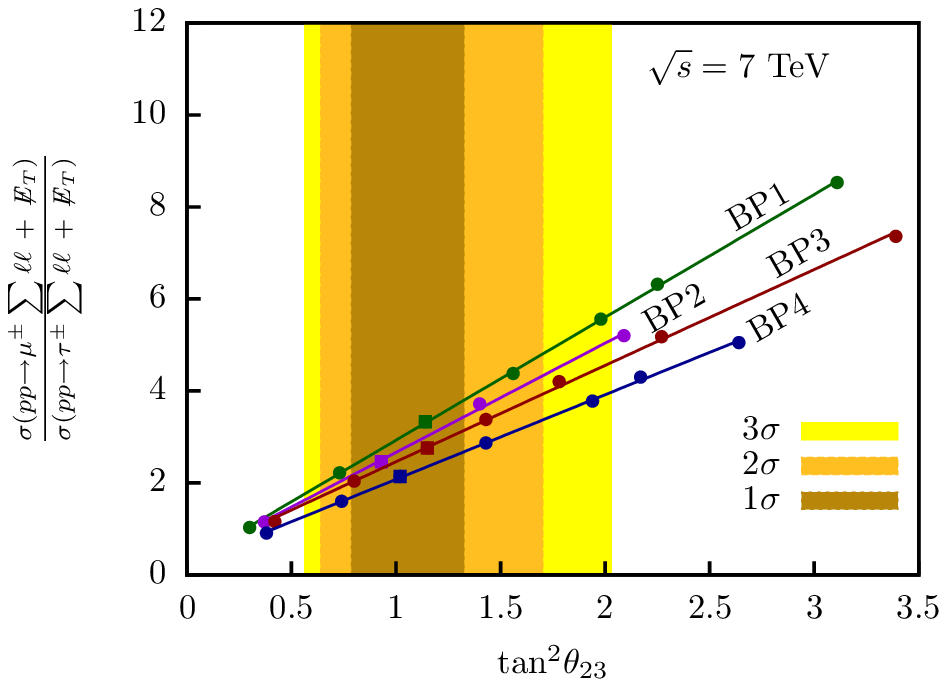,height=7.15cm} 
\caption{Correlation plot ($\frac{\sigma(pp \to \mu^\pm \sum\ell\ell + \mET~)}{\sigma(pp \to \tau^\pm \sum\ell\ell + \mET~)}$
vs $\rm{tan}^2\theta_{23}$, with $\ell=e,\mu$) obtained for LHC at $\sqrt{s}=7~{\rm TeV}$. Three differently
coloured vertical strips correspond to $1\sigma$, $2\sigma$ and $3\sigma$ allowed
region for $\rm{tan}^2\theta_{23}$, respectively. The benchmark points as given in table \ref{input1}, 
table \ref{mass-spectrum} and table \ref{input2} are represented by coloured $\blacksquare$. 
Other points, represented by coloured $\bullet$, are obtained with randomly generated $m^2_{D_i}$ and 
$\delta^2_i$ values.}
\label{correlation_plot1}}
\FIGURE{\epsfig{file=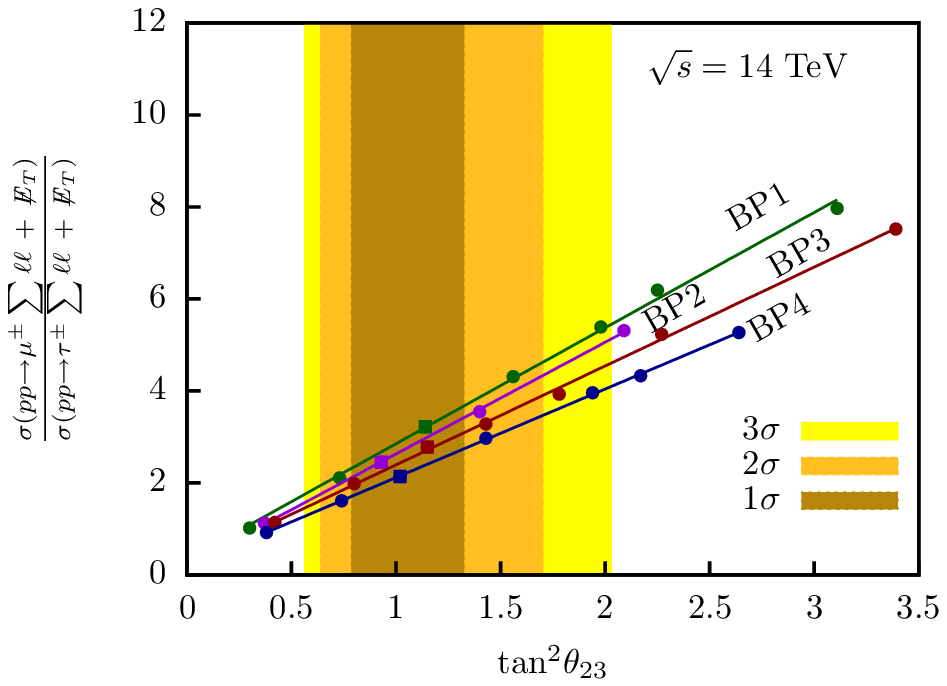,height=7.15cm} 
\caption{Correlation plot ($\frac{\sigma(pp \to \mu^\pm \sum\ell\ell + \mET~)}{\sigma(pp \to \tau^\pm \sum\ell\ell + \mET~)}$
vs $\rm{tan}^2\theta_{23}$, with $\ell=e,\mu$) obtained for LHC at $\sqrt{s}=14~{\rm TeV}$. 
Other specifications are the same as in figure \ref{correlation_plot1}.}
\label{correlation_plot2}}
%
We present these correlations with best fit lines. It can be seen from these figures that the
$3\sigma$ allowed value of $\tan^2\theta_{23}$ \cite{Schwetz:2008er} from atmospheric neutrino oscillation 
experiments predict a value of the ratio of cross sections $\frac{\sigma(pp \to \mu \sum\ell\ell + \mET~)}
{\sigma(pp \to \tau \sum\ell\ell + \mET~)}$, ($\ell = e,\mu$) to be approximately in the range 
$1.0-6$. These predictions can be verified at the LHC or the measured value of this ratio can 
give an alternative estimate of $\tan^2\theta_{23}$. On the other hand, if this ratio comes 
out to be very much different from the ones predicted here then one can perhaps conclude that 
MSISM is not the correct model for explaining neutrino masses and mixing. 

Nevertheless, as we can see, from the correlation plots, that there is a different linear relationship for each 
different kind of benchmark points. In general then, from neutrino oscillation data we cannot give a unique prediction for the ratio of the cross-sections that can be verified at the LHC and help us in constraining the model parameters. In other words, measuring the cross section ratio at the LHC would not allow a prediction of $\theta_{23}$ that could be tested against oscillation results. This means that we need other measurements at the LHC to allow such predictions. As an example, to distinguish among the four benchmark points we plot the ratio,  $m_{\widetilde\chi_1^\pm}/m_{\widetilde N_{1,2}}$ with the ratio of cross-sections of $\mu$ and $\tau$ channels which gives four separate parallel lines for the four benchmark points (figure \ref{bp_separation}).
\FIGURE{\epsfig{file=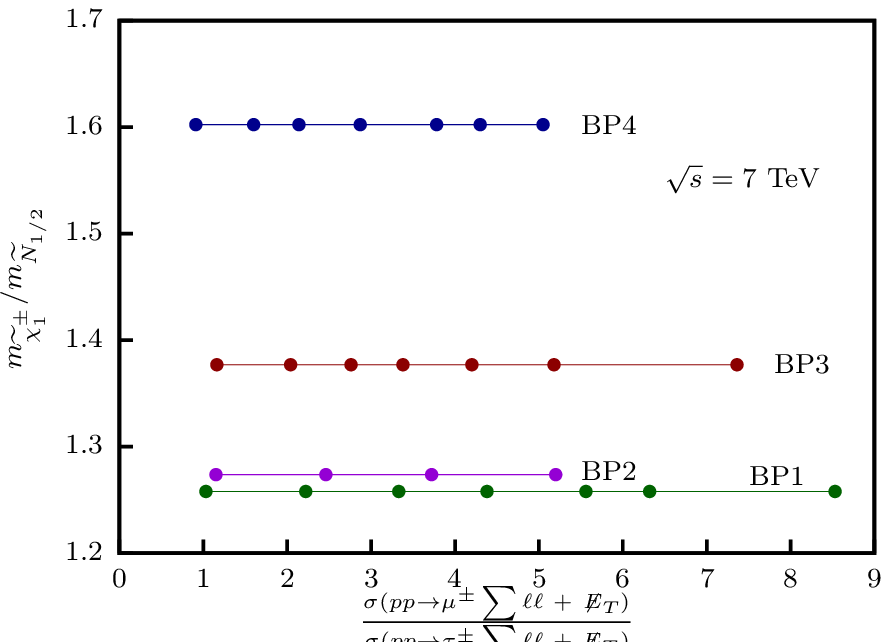,height=6.50cm} 
\vspace*{0.5cm}\caption{${m_{\widetilde\chi_1^\pm}}/{{m_{\widetilde N_{1,2}}}}$ plotted against 
$\frac{\sigma(pp \to \mu^\pm \sum\ell\ell + \mET~)}{\sigma(pp \to \tau^\pm \sum\ell\ell + \mET~)}$ for $\sqrt{s}=7~{\rm TeV}$.
The mass ratios increase as the slopes of the straight lines in figure \ref{correlation_plot1} corresponding to 
the four benchmark points decrease.}
\label{bp_separation}}
One can see from figure \ref{bp_separation} that the ratio (${m_{\widetilde\chi_1^\pm}}/{{m_{\widetilde N_{1,2}}}}$) increases as the slope of the straight lines in the correlation plot corresponding to different benchmark points decreases. This pattern can easily be understood. Increase in the mass ratio indicates greater splitting between the chargino and sneutrino masses. As the splitting increases, the leptons coming from this chargino decay become more energetic (eq.(\ref{signal-choice})). This affects the $\tau$ count in the final state more than the $\mu$ count as the detection efficiency for the taus increases with the increase of visible energy of $\tau$ decay products\cite{Tau-eff}. Hence more $\tau$ events are expected in the final state for those benchmark points which has greater lighter chargino - LSP mass ratio for a given set of $m_{D_i}$'s. With the increase of $\tau$ events the ratio of the cross-sections plotted in the correlation plots decreases and as a consequence gives smaller slope compared to the previous benchmark point. Now it is clearly understood that if we can 
determine the lighter chargino-LSP mass ratio, we can distinguish among the four benchmark points. 

Mass determination techniques in the context of LHC have been studied extensively. Transverse mass variable 
($m_{T_2}$) \cite{Lester:1999tx,Barr:2003rg} is very useful for this purpose. $m_{T_2}$ has also been generalized for the cases where the parent and daughter particles in the two decay chains are not identical \cite{Barr:2009jv,Konar:2009qr}. 
Moreover, final state with more than two invisible particles has also been addressed in ref.\cite{Agashe:2010tu}.
In our case, we observe 
the following: \\
\begin{itemize}
\item The lightest neutralino $\widetilde\chi_1^0$ is also invisible, as mentioned earlier in the text.
\item One lepton is produced from one side of the cascade and remaining two leptons from the other side of the 
cascade (figure \ref{feyn-diag}).
\item $\widetilde\chi_1^\pm$ and $\widetilde\chi_2^0$ are not mass degenerate but the difference is quite
small in the context of mass measurement.
\end{itemize}
So, we see that daughters of different masses are produced here from nearly identical parents. A mass determination 
technique similar to refs. \cite{Barr:2009jv,Konar:2009qr,Agashe:2010tu} can be applied here too to determine 
the masses of the lighter chargino and the sneutrino LSP. 
However, a detailed analysis in this direction is beyond the scope of the present paper. 
Thus we see that measuring the mass ratio ($m_{\widetilde\chi_1^\pm}/{m_{\widetilde N_{1,2}}}$), 
along with the ratio of the trilepton cross-section, can help us pick the correct benchmark point and hence 
predict the correct value of $\theta_{23}$ that could be tested against the oscillation results. On the other hand, 
a precise determination of $\tan^2 \theta_{23}$ from oscillation experiments as well as a measurement of the cross 
section ratio at the LHC can give a unique prediction of the mass spectrum of the model, that can be verified by mass measurements at the LHC.

In support of our explanation for obtaining different slopes, we present the following analysis. Since this difference among the four benchmark points appears because of taking different $\tau$ identification efficiencies for different energy range and for taking separate $p_T$ cuts for $\mu$'s and $\tau$'s, we can remove this by the following strategy: 
\begin{itemize}
 \item A $p_T$ cut of $20~{\rm GeV}$ taken for both $\mu$ and $\tau$.
 \item A uniform $\tau$ identification efficiency of $50\%$ applied over the whole energy range.

\end{itemize}

\FIGURE{\epsfig{file=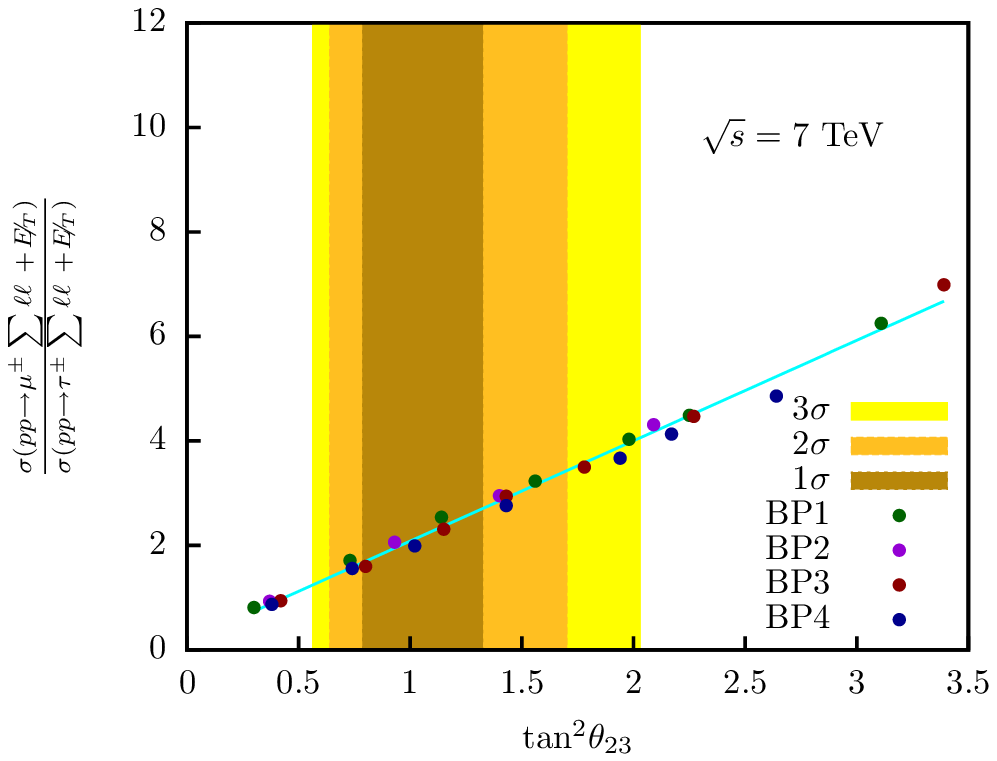,height=7.20cm} 
\vspace*{0.5cm}\caption{Correlation plot ($\frac{\sigma(pp \to \mu^\pm \sum\ell\ell + \mET~)}
{\sigma(pp \to \tau^\pm \sum\ell\ell + \mET~)}$ vs $\rm{tan}^2\theta_{23}$) obtained for $7$ 
TeV center of mass energy under the assumption of uniform $\tau$ identification efficiency ($\epsilon_{\tau}\sim 0.5$) and same $p_T$ cut for both $\mu$ and $\tau$.}
\label{unique_slope}}
We have presented the result in fig.\ref{unique_slope}. This shows the correlation plot for $7~{\rm TeV}$ center of mass energy under the above mentioned conditions. All the benchmark points now lie almost on one straight line. A few points still looks a little bit scattered because of the different isolation criteria used for $\mu$ and $\tau$. 

Finally, in table \ref{ratio_distinct-feature} we show the ratios 
$\frac{\sigma(pp \to e \mu\mu + \mET~)}{\sigma(pp \to 3 \ell + \mET~)}$ 
and $\frac{\sigma(pp \to e e e + \mET~)}{\sigma(pp \to 3 \ell + \mET~)}$ for the four benchmark points.
The smallness of these ratios is also a distinct feature of this model and arises due to the smallness 
of the neutrino reactor angle imposed by neutrino data \cite{Abe:2011sj}. In the usual MSSM scenario 
these ratios are expected to be much higher as there is no suppression of charginos decaying into 
electrons as we have in this model.

\TABLE {\begin{tabular}{|c c c c|} \hline 
&  &$\frac{\sigma(pp \to e \mu\mu + \mET~)}{\sigma(pp \to 3 \ell + \mET~)}\times10^2$
&$\frac{\sigma(pp \to e e e + \mET~)}{\sigma(pp \to 3 \ell + \mET~)}\times10^2$ \\ \hline
     &$\sqrt{s}=7~{\rm TeV}$  & 3.09 &1.06  \\ 
BP1 &$\sqrt{s}=14~{\rm TeV}$ & 3.08 &1.33  \\\hline
     &$\sqrt{s}=7~{\rm TeV}$  & 3.39 &2.13  \\ 
BP2 &$\sqrt{s}=14~{\rm TeV}$ & 3.90 &2.14  \\\hline
     &$\sqrt{s}=7~{\rm TeV}$  & 4.35 &1.71  \\ 
BP3 &$\sqrt{s}=14~{\rm TeV}$ & 4.39 &1.90  \\\hline
    &$\sqrt{s}=7~{\rm TeV}$  & 5.24 &1.86  \\ 
BP4 &$\sqrt{s}=14~{\rm TeV}$ & 5.38 &1.89  \\ \hline
\end{tabular}
\caption{The ratios $\frac{\sigma(pp \to e \mu\mu + \mET~)}{\sigma(pp \to 3 \ell + \mET~)}$ and 
$\frac{\sigma(pp \to e e e + \mET~)}{\sigma(pp \to 3 \ell + \mET~)}$ for the four benchmark points.
\label{ratio_distinct-feature}}}

A discussion of these specific trilepton signals remains incomplete without
a note on the Tevatron analysis of the considered model. For the four 
chosen benchmark points we observed no points with significance $\ge3\sigma$
for the $\tau\sum\ell\ell + \mET~$ final state and simultaneously consistent
with the atmospheric neutrino mixing at the $3\sigma$ limit. This is a well expected result
considering that the Tevatron center-of-mass energy is $1.96$ TeV with $12~{\rm fb^{-1}}$
of integrated luminosity \cite{tev-lumi}. For example, the statistical significance for 
$\tau\sum\ell\ell$ mode for BP1 with $m^2_{D_i}$s given in table \ref{input2} 
is computed to be $1.64$. 
\section{Conclusion}\label{conclusions}
We consider the minimal supersymmetric inverse seesaw model and study its characteristic 
signatures at the LHC. This model, with only one pair of singlet superfields 
explains existing neutrino oscillation data. The model is rich from phenomenological point of 
view and can lead to potentially testable signatures at the hadron colliders. In this R-parity 
conserving model, one of the singlet sneutrino (with a small admixture of the doublet
sneutrino) is the lightest supersymmetric particle (LSP) and as a result shows up in the collider as 
missing energy. Charginos can decay to charged leptons plus singlet sneutrino LSP. The decay patterns 
of the chargino are controlled by the same parameters which generate the neutrino mixing angles. 

In order to study this correlation of the chargino decays and the neutrino mixing angles, we look at 
specific trilepton $+~\mET~$ signatures at the LHC. We show that the ratios of cross sections
of this studied trilepton $+~\mET~$ final states in certain flavour specific channels ($\mu ee + \mET$,
$\mu \mu \mu + \mET$, $\tau ee + \mET$, $\tau \mu \mu + \mET~$) nicely correlate with the atmospheric
neutrino mixing angle. We explore different points in the parameter space to study this correlation. 
A measurement of these cross sections thus provide an interesting test of the minimal supersymmetric 
inverse seesaw model. The hard missing $E_T$ spectrum makes this trilepton final state statistically 
significant by reducing certain standard model background events significantly. We adhere to different 
cuts to reduce the backgrounds coming from some other channels. Motivated by the recent results from 
the ATLAS and the CMS experiments, we work in a scenario with heavy squarks and gluinos and a relatively 
light electroweak sector. The results of our analysis suggest that the theoretical parameter space of this 
model can be constrained by the data collected at the LHC with center-of-mass energy $7$ TeV and for 
an integrated luminosity of $25$ fb$^{-1}$. On the other hand, a measured value of this ratio at the 
LHC can give us an alternative estimate of $\rm{tan}^2 \theta_{23}$ and confirm (or rule out) this minimal 
supersymmetric inverse seesaw model as a possible explanation of neutrino masses and mixing. We also 
show, as a distinct feature of this model, the cross sections of $pp \to e \mu \mu + \mET$ and $pp 
\to eee  + \mET$ are suppressed compared to the total chosen trilepton + $\mET~$ cross section 
because of the restrictions on the neutrino reactor angle imposed by neutrino data.
 

\section*{Acknowledgments}
SM wishes to thank the Department of Science and Technology, Government of India 
for a Senior Research Fellowship.  
SB would like to thank the Department of Theoretical Physics, IACS for the
hospitality while this work was being proposed and initiated. SB would also
like to thank PhD program of HRI and RECAPP, HRI for hospitality and
financial support during the initial phase of this work.
PG would like to thank the Council of Scientific and Industrial Research, 
Government of India for the financial assistance during the initial phase of this work.
PG's work is supported by the Spanish MICINN under grant FPA2009-08958. PG also thank the 
support of the MICINN under the Consolider-Ingenio 2010 Programme with grant MultiDark 
CSD2009-00064, the Community of Madrid under grant HEPHACOS S2009/ESP-1473, and the European 
Union under the Marie Curie-ITN program PITN-GA-2009-237920. PG and SM wish to thank RECAPP, 
HRI for hospitality during a part of the investigation. We are very grateful to Asesh Krishna 
Datta for some very helpful discussions. We also thank Nabanita Bhattacharyya, Anindya Datta, 
Martin Hirsch, Partha Konar, Satyanarayan Mukhopadhyay and Sujoy Poddar for many useful comments 
and suggestions. PG is grateful to Chan Beom Park for his insightful suggestions.
SR acknowledges the hospitality provided by the Helsinki Institute of Physics 
and the CERN Theory Group during the final phase of this work. 
\appendix

\bibliography{invseesaw_LHC_v11}
\end{document}